\documentclass[preprint,3p,sort&compress]{elsarticle}
\usepackage{amsmath,amssymb}
\usepackage{epsfig}
\usepackage{color}
\usepackage{verbatim}
\usepackage[normalem]{ulem}
\usepackage[utf8]{inputenc}

\def\gs{\mathrel{
   \rlap{\raise 0.511ex \hbox{$>$}}{\lower 0.511ex \hbox{$\sim$}}}}
\def\ls{\mathrel{
   \rlap{\raise 0.511ex \hbox{$<$}}{\lower 0.511ex \hbox{$\sim$}}}}

\newcommand{\ba}{\begin{array}{c}}
\newcommand{\baz}{\begin{array}{cc}}
\newcommand{\bad}{\begin{array}{ccc}}
\newcommand{\bav}{\begin{array}{cccc}}
\newcommand{\baf}{\begin{array}{ccccc}}
\newcommand{\bena}{\begin{eqnarray}}
\newcommand{\eena}{\end{eqnarray}}
\newcommand{\bea}{\begin{equation} \begin{array}{c}}
\newcommand{\eea}{ \end{array} \end{equation}}
\newcommand{\ea}{\end{array}}

\journal{Astroparticle Physics}

\usepackage{amsmath,amssymb}

\begin{document}

\title{Neutrinos from the gamma-ray source eHWC J1825-134:\\ predictions for Km$^3$ detectors}

\author[apc]{V. Niro}
\ead{viviana.niro@apc.in2p3.fr}
\author[apc,switz]{A. Neronov}
\ead{andrii.neronov@apc.in2p3.fr}
\author[apc]{L. Fusco}
\ead{luigi.fusco@apc.in2p3.fr} 
\author[apc]{S. Gabici}
\ead{gabici@apc.in2p3.fr}
\author[apc]{D. Semikoz}
\ead{semikoz@apc.in2p3.fr}

\address[apc]{APC, AstroParticule et Cosmologie, Universit\'e Paris Diderot, CNRS/IN2P3, CEA/Irfu, 
Observatoire de Paris, Sorbonne Paris Cit\'e, 10, rue Alice Domon et L\'eonie Duquet, 75205 Paris Cedex 13, France} 
\address[switz]{Astronomy Department, University of Geneva, Ch. d’Ecogia 16, 1290, Versoix, Switzerland}

\begin{abstract}
The eHWC J1825-134 source is located in the southern sky and has been 
recently detected by the HAWC observatory. It presents an hard spectral 
index and its gamma-ray flux extends up to energies close to 100 TeV without significant suppression. 
Amongst the HAWC sources, it is the most luminous in the multi-TeV domain and therefore is one of the 
first that should be searched for with a neutrino telescope in the 
northern hemisphere. 
Using an updated effective area for the forthcoming KM3NeT detector, 
we study the possibility to detect this source within ten years of its 
running time. 
We conclude that about a 4 to 5 sigma detection has to be expected after ten years of observations, depending on the details of the considered scenario. 
\end{abstract}


\maketitle

%%%%%%%%%%%%%%%%%%%%%%%%%%%%%%%%%%%%%%%%%%%%%%%%%%%%%%%%%%%%%%%%%%%%%%
\section{\label{sec:intro} Introduction}
%%%%%%%%%%%%%%%%%%%%%%%%%%%%%%%%%%%%%%%%%%%%%%%%%%%%%%%%%%%%%%%%%%%%%%
The observed energy spectrum of cosmic rays is described by a power law with spectral index of about 2.7 up to energies of a few PeV, where the spectrum gets steeper and a feature called the ``knee'' originates~\cite{Gabici:2019jvz,Kachelriess:2019oqu}. 
The knee is believed to mark the maximum energy for cosmic rays accelerated at Galactic sources \cite{2006JPhCS..47..168H},  or alternatively the energy above which the effectiveness of the confinement within the Galaxy is reduced \cite{2014PhRvD..90d1302G}.

The problem of the origin of galactic cosmic rays is one of the most important problems in high energy astrophysics \cite{Gabici:2019jvz,Kachelriess:2019oqu}. This is particularly true for energies around the knee or greater, since explaining the origin of cosmic rays in that energy range is problematic~\cite{Bell:2013kq}. Different possible sources of galactic cosmic rays have been proposed, among which supernova remnants, proposed in 1934 by Baade and Zwicky, are the most accredited ones~\cite{Drury:2017sna}. It is, however, not clear whether supernova remnants can accelerate cosmic rays up to PeV energies or if other sources should be considered. 
With this respect, we recall that evidence for the acceleration of PeV particles in the Galactic centre has been reported by the H.E.S.S. Collaboration~\cite{Abramowski:2016mir}.
 
During the acceleration of cosmic rays, the production of gamma rays is expected. These could be produced from the decay of neutral pions, arising from the hadronic interactions with the interstellar medium, or from leptonic processes, like inverse compton, that is however suppressed by Klein-Nishina effect in the multi-TeV energy domain \cite{:2007qGabicib}. 
The identification of the origin of the gamma-ray emission, specifically if it is leptonic or hadronic is thus one of the most important goals in gamma-ray astronomy. 

If cosmic rays lose part of their energy in hadronic processes, then, a flux of high energy neutrinos is expected from charged pion decays. Neutrino telescopes are for this reason able to provide important information on the production mechanisms of cosmic rays as the detection of neutrinos from a gamma-ray source would imply that the emission is hadronic. 

From the data collected in 7.5 years of running of the IceCube detector, 103 neutrino events were identified, of which 60 events with deposited energy $E_{dep}>$ 60 TeV~\cite{Schneider:2019ayi}. 

At present, the event distribution is consistent with isotropy and therefore often interpreted in terms of extragalactic sources, see for example Ref.~\cite{Ahlers:2018dtq} for a recent review. 
However, also Galactic scenarios have been proposed,
see for example Refs.~\cite{Taylor:2014hya,Neronov:2018ibl} on this topic. 
On the galactic and extragalactic contributions of the flux see also the analyses in Refs.~\cite{Neronov:2015osa,Neronov:2013lza,Palladino:2016zoe,Pagliaroli:2016lgg,Gaggero:2015xza,Ahlers:2015moa}.
Moreover, at the moment a 3.5$\sigma$ evidence is present for neutrino emission coming from the
direction of the blazar TXS 0506+056, see Refs.~\cite{IceCube:2018cha,Padovani:2018acg,Gao:2018mnu,Keivani:2018rnh,Murase:2018iyl}. 

A multi-messenger search is mandatory for the identification of the origin of cosmic neutrinos. Indeed, gamma-ray data are necessary to make correct estimation of neutrino fluxes from point-sources. The characteristic gamma-ray feature of a PeVatron include an hadronic, hard spectrum that extends until at least several tens of TeV. 
To search for these PeVatrons a gamma-ray experiment with detection sensitivity up to about 100 TeV is of fundamental importance. 

The High Altitude Water Cherenkov Observatory (HAWC) is a gamma-ray observatory  sensitive in the multi-TeV energy domain. For this reason, it is currently 
the most sensitive gamma-ray detector for discovering PeVatrons. 
The HAWC observatory has reported new data on galactic sources in recent years, see e.g.~\cite{Abeysekara:2017hyn,Malone:2018zeg,Abeysekara:2019gov}. Among these sources, the eHWC J1825-134 source, located in the southern sky, has been detected with an hard spectrum that extends up to multi-TeV energies, thus it represents a possible PeVatron source. Moreover, this is the brightest source detected by HAWC in the multi-TeV domain. 

Note that the IceCube detector has an optimal sensitivity for sources located in the northern hemisphere, and is less sensitive to sources located in the southern sky, 
using tracks events. It is roughly an order of magnitude less sensitive if one considers only track events and a spectrum of the type $dN/dE \propto E^{-2}$, while more than two orders of magnitude for a source $dN/dE \propto E^{-3}$, see Fig.~3 in Ref.~\cite{Aartsen:2019epb}. The use of cascades events, arising from neutrino of all flavours, and of the DeepCore sub-arrays improve the sensitivity 
of IceCube to the sources in the southern celestial hemisphere, compared to the
use of only tracks events~\cite{Aartsen:2019epb}. For a search of several TeV gamma-ray sources observed by H.E.S.S. in the southern sky with IceCube we refer to Ref.~\cite{Aartsen:2019sid}. 
A kilometer-cube detector in the northern hemisphere, instead, will see these events as muon events, for which a good angular reconstruction is possible and could use all its volume for the point sources analysis. 
The importance of a kilometer-cube detector in the northern hemisphere was pointed out considering numerous galactic sources in Ref.~\cite{Kistler:2006hp}. Moreover, it was previously considered also in connection with the prospects of detecting young supernova remnants~\cite{Vissani:2011vg,Vissani:2011ea} and the Milagro diffuse flux from the inner galaxy~\cite{Gabici:2008gw}.

Several studies have been carried out about the possible detection of Galactic sources in the northern hemisphere at IceCube, in particular considering sources detected by the Milagro Collaboration, see e.g.~\cite{Halzen:2008zj,Kappes:2009zza,GonzalezGarcia:2009jc}. 
It was found that for specific sources, a discovery at 3 standard deviation in less than 10 years is feasible~\cite{Gonzalez-Garcia:2013iha,Halzen:2016seh}.
However, the predictions are affected by the discrepancies between information coming from different gamma-ray experiments, 
air-Cherenkov telescopes and air-shower detectors, probably due to the different field of view and energy range. For a recent 
update, about the Milagro sources and considering the new HAWC data, 
we refer to Ref.~\cite{Kheirandish:2019bke}. 

In this work we present modelling of the source eHWC J1825-134. 
In Section~\ref{sec:HAWC}, we describe the data present in the literature on the source, while in 
Section~\ref{sec:aeff}, the calculation of the neutrino flux and the KM3NeT/ARCA effective area is considered. 
In Section~\ref{sec:res}, we present our results, while in Section~\ref{sec:conclusion} our conclusions.

%%%%%%%%%%%%%%%%%%%%%%%%%%%%%%%%%%%%%%%%%%%%%%%%%%%%%%%%%%%%%%%%%%%%%%
\section{\label{sec:HAWC} The eHWC J1825-134 source}
%%%%%%%%%%%%%%%%%%%%%%%%%%%%%%%%%%%%%%%%%%%%%%%%%%%%%%%%%%%%%%%%%%%%%% 
As motivated in the introduction, in this work we will consider the source eHWC J1825-134, analysed in Ref.~\cite{Abeysekara:2019gov}. 
This source is located in the southern sky with a right ascension of $276.40^\circ$ and 
a declination of $-13.37^\circ$. 
We will use for the analysis the spectrum reported in Ref.~\cite{Abeysekara:2019gov}, where a  power law with exponential cut-off fit was considered: 
\begin{equation}
\frac{dN_\gamma}{dE_\gamma} = \phi_0~\left(\frac{E_\gamma}{10~{\rm TeV}}\right)^{-\alpha_\gamma}~\exp{\left(-\frac{E_\gamma}{E_{cut,_\gamma}}\right)}
\end{equation}
with $E_{cut,_\gamma}$ being the cut-off energy of the gamma-ray spectrum, $\alpha_\gamma$ the spectral 
index and $\phi_0$ the flux normalized, see values in Table~\ref{tab:sources_fit}. \\
\begin{table}[!h]
\centering
\begin{tabular}{l || c | c | c | c }
\hline
Source & $\sigma_{ext}$ & $\phi_0$ & $\alpha_\gamma$ & $E_{cut,\gamma}$\\
[1ex]\hline \hline
eHWC J1825-134 & $0.53\pm0.02$ & $2.12\pm0.15$ & $2.12\pm 0.06$ & $61\pm12$ \\[0.5ex]
\hline
\end{tabular}
\caption{Extension of the source in degrees, flux $\phi_0$ in units of $10^{-13}~{\rm TeV}^{-1}~{\rm cm}^{-2}~{\rm s}^{-1}$, spectral index $\alpha_\gamma$ and cut-off energy $E_{cut, \gamma}$. 
}
\label{tab:sources_fit}
\end{table}
The sensitivity of HAWC to the high energy tail of the spectrum is of 
fundamental importance for the correct prediction of the neutrino flux. 
Note that the flux provided by HAWC for this source at 10~TeV is higher than the one reported by the same collaboration for the 
Crab nebulae~\cite{Abeysekara:2017mjj,Abeysekara:2019edl}, that is about $10^{-13}$~TeV$^{-1}$~cm$^{-2}$~s$^{-1}$ considering the one sigma systematic error. 
Thus, this source is one of the brightest sources in the sky and one of the first that should be considered in the analysis of neutrino flux for the KM3NeT detector. 

Finally, it should be noted that the region under examination is quite complex. First of all, as pointed out in \cite{Abeysekara:2019gov}, eHWC J1825-134 overlaps with two HESS sources: the very bright HESS J1825-137 \cite{Aharonian:2006zb,Abdalla:2018qgt} and the much weaker HESS J1826-130 \cite{2017arXiv170804844A}. Second, as we will show in Sec.~\ref{sec:res}, Fermi/LAT data reveal the presence of an extended emission in the region.

%%%%%%%%%%%%%%%%%%%%%%%%%%%%%%%%%%%%%%%%%%%%%%%%%%%%%%%%%%%%%%%%%%%%%%
\section{\label{sec:aeff} The neutrino flux and the KM3NeT/ARCA effective area}
%%%%%%%%%%%%%%%%%%%%%%%%%%%%%%%%%%%%%%%%%%%%%%%%%%%%%%%%%%%%%%%%%%%%%%
In this work we will consider 
the possible detection of the source eHWC J1825-134 at the KM3NeT detector through tracks events originated by muon neutrino charged current interactions. 

In this section, we report the calculation for the neutrino
events, using the KM3NeT effective area. In particular, we considered
the Letter of Intent of the KM3NeT collaboration which contains the
expected performance of the KM3NeT/ARCA
detector~\cite{Adrian-Martinez:2016fdl}. 
Note that the effective area for muon neutrinos has to be corrected by the background
rejection efficiency, in order to account for the loss in events due to
event selections. In order to obtain an approximate value for this procedure, the
information contained in the Letter of Intent of the KM3NeT collaboration is
used. For this specific case, we will consider the selection cut in the
parameter $\Lambda$ reported there, which gives an effective area
optimized for energies above~1~TeV. The total effective area, the
selection efficiency, and the effective area optimized for energies
above~1~TeV are reported in the left panel of 
Fig.~\ref{fig:neutrino and muon neutrino flux}.

The event rate at KM3NeT can be described by 
the expression reported in Ref.~\cite{Kappes:2006fg}:
\begin{eqnarray}
N_{\rm ev}= \epsilon_\theta \epsilon_v\,t\, \int_{E_\nu^{\rm th}} dE_\nu ~\frac{dN_\nu(E_\nu)}{dE_\nu} \times A_\nu^{\rm eff}\,, 
\label{eq:nevmus}
\end{eqnarray}
where a sum over neutrino and antineutrino contributions is implicit. The parameter $\epsilon_v=0.57$ is the visibility of the source, while $\epsilon_\theta=0.72$ takes into account a reduction factor due to the fact that only a fraction of the signal will be detected if the source morphology is assumed to be a Gaussian of standard deviation $\sigma_{ext}$ and the signal is extracted within a circular region of radius $\sigma_{\rm eff} = 1.6 \sqrt{\sigma_{\rm ext}^2+\sigma_{\rm res}^2}$ \cite{Alexandreas:1992ek}. 
Here, $\sigma_{\rm res} \sim 0.1^{\circ}$ is the angular resolution of KM3NeT/ARCA \cite{Adrian-Martinez:2016fdl}.
The number of neutrino events $\frac{dN_\nu(E_\nu)}{dE_\nu}$ has been calculated starting from the gamma-ray 
spectrum and considering the expressions given in Ref.~\cite{Villante:2008qg}, see also Ref.~\cite{Kelner:2006tc} for another derivation. 

The expected atmospheric muon neutrinos are calculated as described in Ref.~\cite{Gonzalez-Garcia:2013iha}, using Refs.~\cite{Honda:2011nf,Volkova:1980sw,Gondolo:1995fq}. 
The flux is then integrated over an opening angle equal to 
$\Omega=\pi \sigma_{\rm eff}^2$.

%%%%%%%%%%%%%%%%%%%%%%%%%%%%%%%%%%%%%%%%%%%%%%%%%%%%%%%%%%%%%%%%%%%%%%
\section{\label{sec:res} Results}
%%%%%%%%%%%%%%%%%%%%%%%%%%%%%%%%%%%%%%%%%%%%%%%%%%%%%%%%%%%%%%%%%%%%%%

\subsection{Fermi/LAT observations}

\begin{figure}[!t]
\centering
\begin{tabular}{rl}
\includegraphics[width=\textwidth]{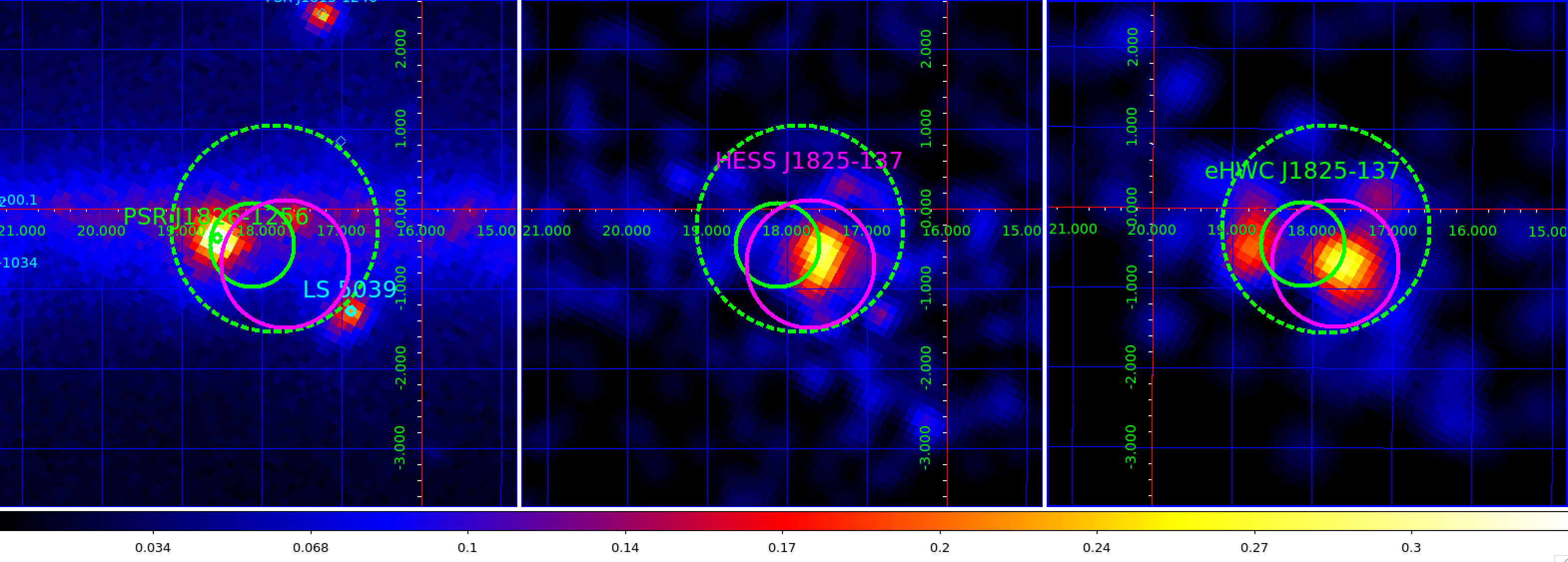} 
\end{tabular}
\caption{\label{fig:hess} 
Fermi/LAT count maps of the source region in 1-10, 100-300 and $>300$ GeV energy ranges (left to right). The 1-10 GeV and 100-300 GeV maps are smoothed with 0.3 degree Gaussian, the $300$~GeV map is smoothed with 0.5 degree Gaussian.}
\end{figure}

\begin{figure}[!t]
\centering
\begin{tabular}{rl}
\includegraphics[width=0.8\textwidth]{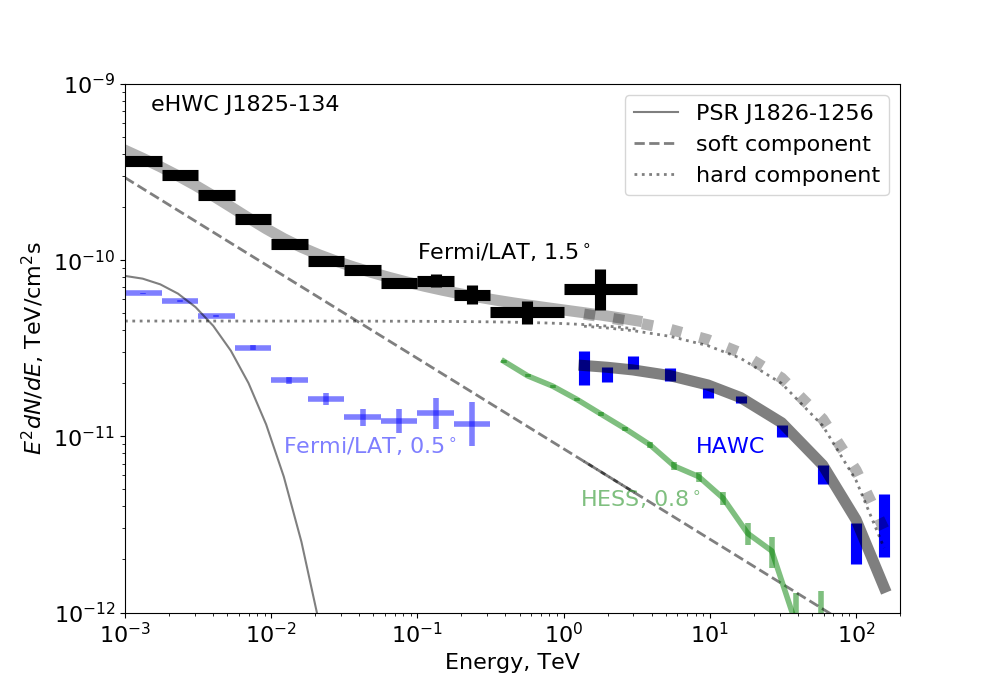} 
\end{tabular}
\caption{\label{fig:spectrum} 
Spectrum of eHWC J1825-134 region measured by Fermi/LAT compared to the HAWC and HESS spectral measurements. Blue thin data points show the spectrum extracted from a circular region of the radius $0.5^\circ$ 
around the source position of the HAWC source. Black thick data points are for the spectrum extracted from the $1.5^\circ$ radius region shown in 
the right panel of Fig. \ref{fig:hess}. Dashed, dotted and thin solid curves show spectral fit components. 
The hard component (dotted line) shape is adjusted to fit the HAWC data above 1 TeV. Its normalization is found from the fit to the Fermi/LAT data.}
\end{figure}

Before proceeding with the estimate of the expected neutrino flux, we report in Figs.~\ref{fig:hess},~\ref{fig:spectrum} the results of our 
analysis of the region using Fermi/LAT data. For this analysis we have used events of the {\tt SOURCEVETO} class which are characterised by low residual cosmic ray background 
contamination~\cite{Bruel:2018lac,Neronov:2019ncc}. We have filtered the events collected within time interval $246758401$~s$<$MET$<582686231$~s using the {\it gtselect-gtmktime} sequence as described in Fermi/LAT analysis threads\footnote{https://fermi.gsfc.nasa.gov/ssc/data/analysis/scitools/}. Fig.~\ref{fig:hess} shows the count maps of the source region in different energy ranges. The left panel shows the 1-10 GeV map smoothed with 0.3 degree Gaussian. The dominant source in the region is the pulsar PSR J1826-1256. The HAWC source (green solid circle) is immediately adjacent to the pulsar location. The pulsar is not visible in the energy range above 100 GeV, as one could see from the middle panel of Fig.~\ref{fig:hess}. In this energy range the centroid of the source is at the position of the extended source HESS J1825-137, identified as a pulsar wind nebula. In the energy range above 300 TeV the source centroid shifts back toward the position of the pulsar. This might explain the mismatch between the source positions measured by HESS and HAWC. One could notice that the Fermi/LAT source consists of two components and the position of the HAWC source is in between them. 
Therefore, given this complicated source morphology, it is not possible to find an exact match between the HAWC extended source and different source components observed by Fermi/LAT. 

In Fig.~\ref{fig:spectrum} we show the results of the Fermi/LAT spectral analysis which is based on the aperture photometry approach (the only viable approach given the uncertain morphology of the source). Blue thin data points show the spectrum of the source region extracted from $0.5^\circ$ radius circle centered at the position of the HAWC source (green solid circle). How does this compare with the flux measured by HAWC?

The total flux of the HAWC source (blue data points in Fig.~\ref{fig:spectrum}) has been extracted within a large region, under the assumption that the source has a Gaussian morphology. The region containing 68\% of the HAWC flux has a radius of $\sigma_{ext} = 0.53$ degrees \cite{Abeysekara:2019gov} and is indicated with the green solid circle in Fig.~\ref{fig:hess}.
Therefore, one can see that the HAWC flux within $\sigma_{ext}$ is slightly larger than the flux measured by Fermi/LAT, but still consistent within the statistical errors. 
We report in the Figure also the data points for HESS J1825-137 considering an opening angle of 0.8$^\circ$~\cite{Abdalla:2018qgt}. 

The black data points in Fig.~\ref{fig:spectrum} show the source spectrum extracted from the region which encompasses the $\sim$~1~TeV emission as observed by Fermi/LAT 
(the green dashed circle of radius $1.5^\circ$ in Fig. \ref{fig:hess}), we find that the flux level measured by Fermi/LAT in the TeV range is somewhat higher than that of HAWC. 
Note that the HAWC analysis assumes a Gaussian source morphology convolved with the HAWC point spread function, which does not match the complex 
morphology seen by Fermi/LAT.

The $1.5^\circ$ region of the HAWC/HESS source includes, apart from the extended source itself, also the pulsar PSR J1826-1256 and the diffuse emission from the Galactic disk in front/behind the HAWC/HESS source. Taking this into account, we model the source spectrum measured by Fermi/LAT with three model components. For the PSR J1826-1256, we adopt the spectrum cited in the Fermi 4FGL catalog \cite{Fermi-LAT:2019yla}. The spectrum of the diffuse emission from the inner Galactic disk is well modelled with 
the power law spectrum with the slope $\Gamma\simeq 2.5$ \cite{Neronov:2015vua,yang_aharonian}. The spectrum shown in Fig. \ref{fig:spectrum} exhibits a high-energy hardening, which could be modelled adding a 
cut-off power law component to the spectral model. Fitting together the sum of the soft power law (Galactic Disk), the hard cut-off power law (the HAWC / HESS source), and the pulsar, we find the fit shown in Fig.~\ref{fig:spectrum}. 
We find that the normalisation of the cut-off power law found from the fit is by factor of 1.5  higher for the Fermi/LAT as compared to the HAWC spectral fit. 

The discrepancy between the Fermi/LAT and HAWC fluxes at photon energy around 1 TeV could be possibly ascribed either to the simple source morphology assumed to extract the HAWC flux,
or to the difficulty of estimation of the cosmic ray background in the source region \cite{Neronov:2019ncc}, or to systematic errors \cite{Abeysekara:2019gov}. 
Given these uncertainties, in the following we estimate the expected number of neutrino events in KM3NeT from the region in two different scenarios.

\subsection{Neutrino event rate from the eHWC J1825-134 source}

\begin{figure}[!t]
\centering
\begin{tabular}{rl}
\includegraphics[width=0.48\textwidth]{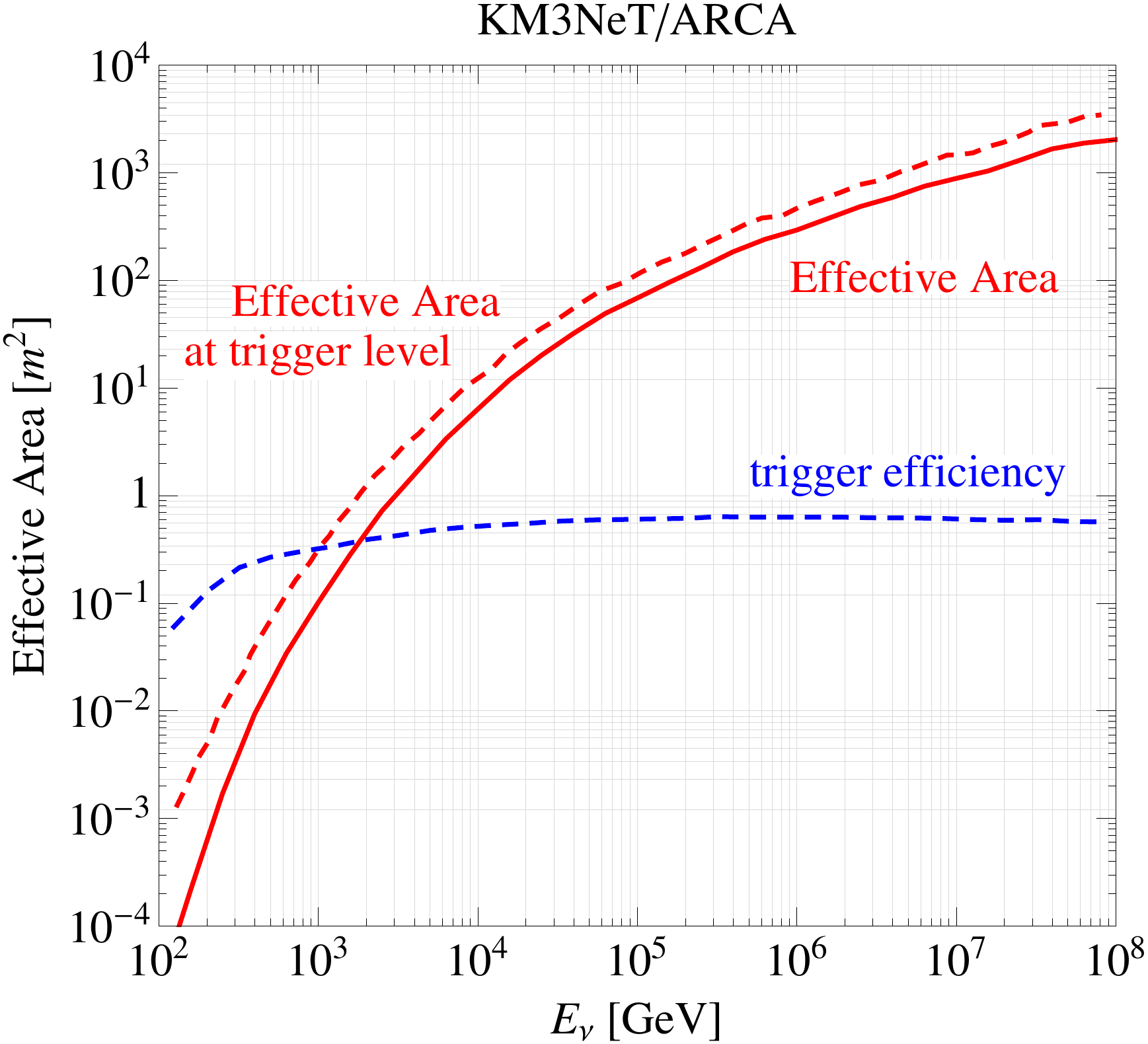} &
\includegraphics[width=0.48\textwidth]{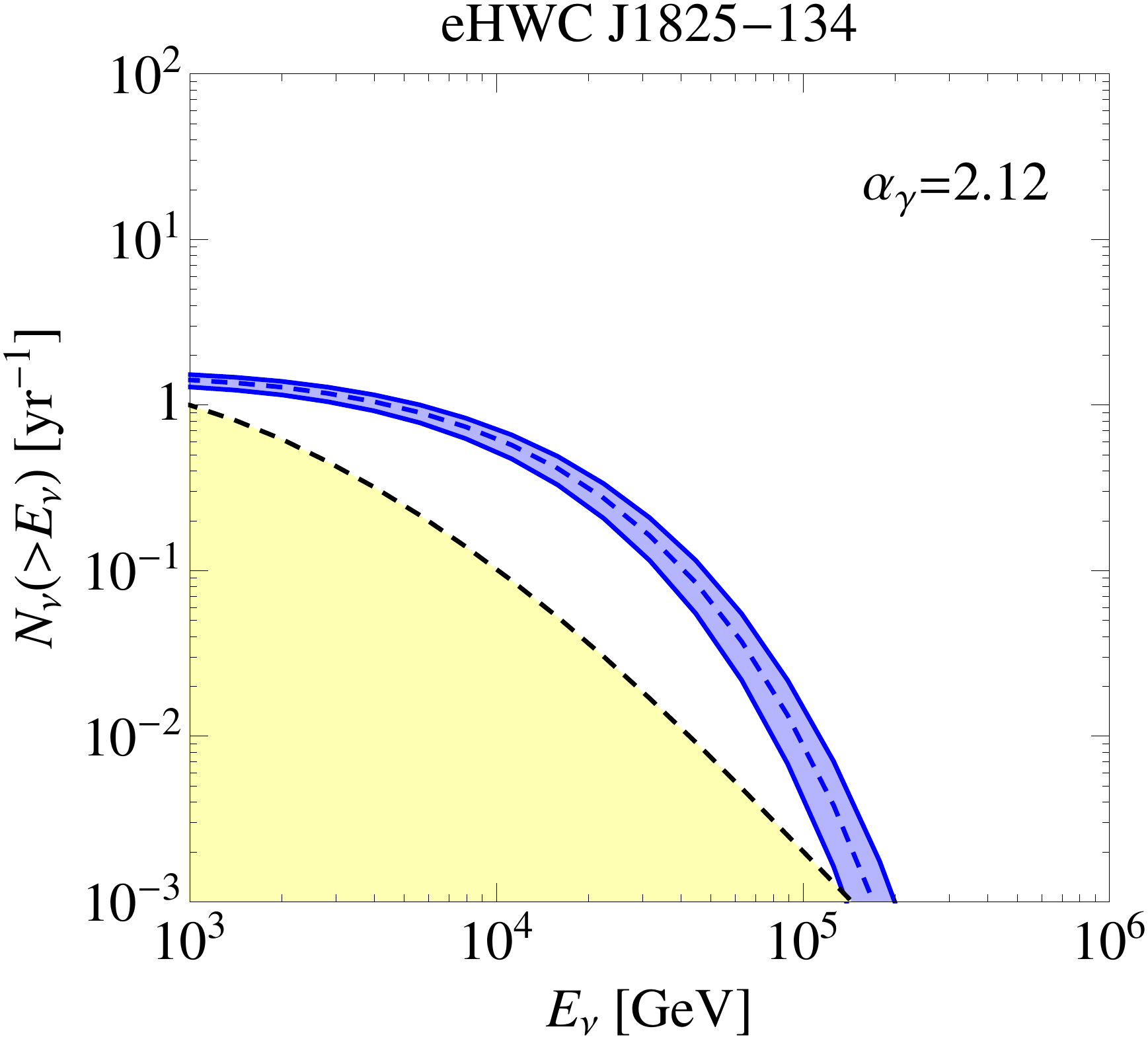}
\end{tabular}
\caption{\label{fig:neutrino and muon neutrino flux} 
Right: We show the effective area used in the analysis (red solid line), the effective area at trigger level (red dashed line), and the trigger efficiency (blue dashed), which gives an effective area optimized for energies above 1~TeV; Left: number of events expected for 
the atmospheric background (yellow area) and for the source for the best-fit value of $\alpha_\gamma$ and different values of $E_{cut,\gamma}$. The blue band 
represents the statistical errors in $E_{cut,\gamma}$.}
\end{figure}

In this section we estimate the neutrino flux from the HAWC source as described in Section \ref{sec:aeff}. We have fixed 
the normalization and the size of the source to its best-fit values. 
The spectral index has been also fixed to the best-fit value $\alpha_\gamma=2.12$. 
The energy of the cut-off, instead, has been varied within the statistical errors. The results are reported in the right panel of Fig.~\ref{fig:neutrino and muon neutrino flux}. 
We report in Table~\ref{tab:events}, the number of atmospheric 
neutrino events $N_{\rm atm}$ and the number of source events $N_{\rm src}$, above the 
following neutrino energy threshold: $E_\nu^{\rm thr}>$ 1,~10,~30 and 100~TeV, for 10 years running time of the KM3NeT detector. Note that since we are considering an 
effective area optimized for energies above~1~TeV, we chose as lower energy threshold 
1~TeV. The other values are reported to show the number of events in case of 
analyses optimized for higher energies. 
As it is clear from the table, the signal events are always significantly above the background as long as the energy threshold is below~10~TeV. 
If we consider an energy threshold of about 30~TeV the signal events are reduced to 1.8, in case of $E_{\rm cut, \gamma} =61$~TeV, while to 1.3 and 
2.3 in case of $E_{\rm cut, \gamma}$ within the statistical error band. For even higher energy threshold of about 100~TeV, the number of expected signal events is below 1. 
We have estimated the statistical significance as reported in Ref.~\cite{ATLAS:2011tau} 
and as described in Refs.~\cite{Halzen:2016seh,Gonzalez-Garcia:2013iha}. 
We report in the left panel of Fig.~\ref{fig:p_value} the results for the p-value as a function of the energy threshold for 10 years of running time of the KM3NeT detector and $\alpha \sim 2.12$. We can see that for an energy threshold of the order of $E_\nu^{\rm thr} \lesssim 10$~TeV we have a minimum in the p-value. 
We can see that in 10 years of running of 
KM3NeT the significance is well above 3$\sigma$ as long as the energy threshold is less than about 10~TeV. 

\subsection{Neutrino event rate from the eHWC J1825-134 extended region}

Here, we estimate the neutrino event rate in KM3NeT considering the region of 1.5$^{\circ}$ radius indicated as a dashed green circle in Fig.~\ref{fig:hess}. We take the Fermi/LAT flux as reference, and so we repeat the calculation as in Sec.~\ref{sec:aeff} by multiplying the HAWC gamma-ray flux by a factor of 1.5. In order to account for the more extended region, we set $\sigma_{\rm eff} = 1.5^{\circ}$, and moreover $\epsilon_{\theta} = 1$ (i.e. we do not assume a Gaussian morphology).

\begin{figure}[!t]
\centering
\begin{tabular}{rl}
\includegraphics[width=0.48\textwidth]{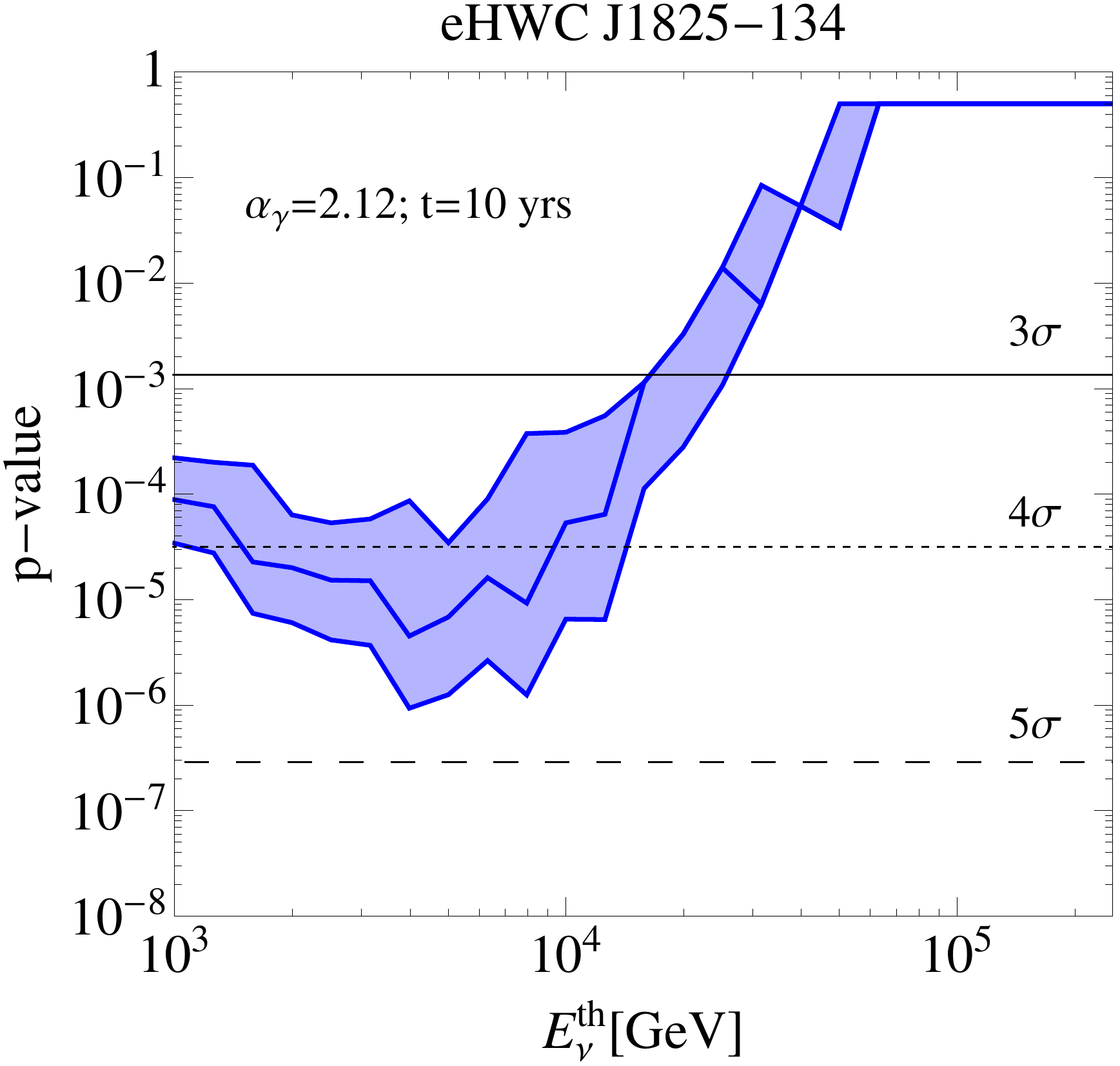} & 
\includegraphics[width=0.48\textwidth]{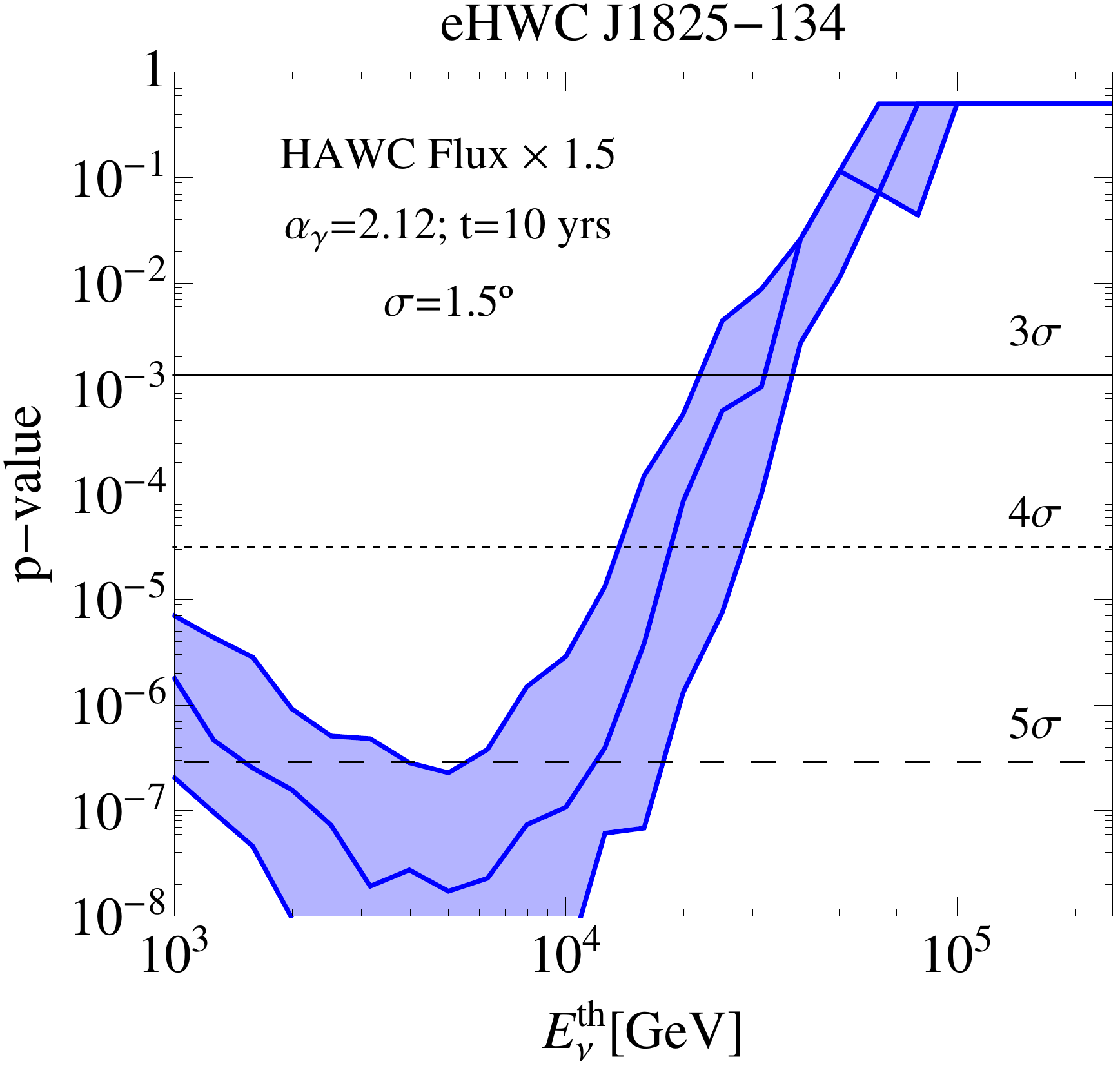} 
\end{tabular}
\caption{\label{fig:p_value} 
p-value for the best-fit value of $\alpha_\gamma$ and different values of $E_{cut,\gamma}$ for 10 years of running of the KM3NeT detector. The blue band 
represents the statistical errors in $E_{cut,\gamma}$. In the left panel, we have considered the normalization best-fit as provided by the HAWC collaboration and the extension of the source is fixed to 0.53$^\circ$, while in the 
right panel we have considered the HAWC flux multiplied by 1.5 and an opening angle of $1.5^\circ$. }
\end{figure}

The results are shown in the right panel of Fig.~\ref{fig:p_value}, where the p-value is plotted. In this case, the statistical 
significance reaches $5\sigma$ in the case of 10 years running time and for an energy threshold of the 
order of about 10~TeV. 

\subsection{Comparison with previous studies}

This source was previously studied in Ref.~\cite{Ambrogi:2018skq}, where an effective area with 6 building blocks for the 
KM3NeT detector was considered and an angular opening of 0.9$^\circ$ radius. 
The authors found that the neutrino flux from the source 2HWC J1825-134 is above the sensitivity for 
10 years running of the KM3NeT detector within a wide range of energies. 

Comparing with the KM3NeT study for point-sources with a spectrum 
$dN/dE \propto E^{-2}$, we can see that a source with the same normalization of 
eHWC J1825-134 is above the sensitivity for a $5\sigma$ detection at the KM3NeT detector within 3 years of running time~\cite{Adrian-Martinez:2016fdl}. 
Considering the cut-off energy, a discovery at more than $4\sigma$ could be reached into about 10 years. 

This source is below the sensitivity of ANTARES data reported in 
Ref.~\cite{Adrian-Martinez:2016fdl} for point-sources with a spectrum 
$dN/dE \propto E^{-2}$, see also Ref.~\cite{Albert:2017ohr} 
for the ANTARES 7 years of data for a source with $E_{cut,\nu}<100$~TeV. 

For the possibility of discovering this source at the IceCube 
detector, we refer to Ref.~\cite{Aartsen:2019epb}, where an analysis 
for the 7-years tracks and cascade events was considered for specific value of the cut-off energy $E_{cut,\nu}=100$~TeV and 1~PeV. 
From what reported, the cascade channels represent the most promising way to discover this source at the IceCube detector.

\begin{table}
\centering
 \begin{tabular}{c || c c c c } 
 \hline
{\it Events in 10 yrs} & $E_\nu^{\rm thr}> 1~{\rm TeV}$ & $> 10~{\rm TeV}$ & $> 30~{\rm TeV}$ & $> 100~{\rm TeV}$ \\ [1ex] 
 \hline\hline
$N_{\rm atm}$
& 10.0 
& 1.1
& 0.2 
& 0.02 
\\ [1ex]\hline 
$N_{\rm src}${\it (best-fit)} 
& 14.2
& 6.3 
& 1.8 
& 0.1 \\ [1ex]\hline
$N_{\rm src}${\it (statistical)} 
& 12.8;~15.3 
& 5.2;~7.2 
& 1.3;~2.3 
& 0.06;~0.2\\ [1ex] \hline 
\end{tabular}
\caption{\label{tab:events}Number of events for the atmospheric background, $N_{\rm atm}$, and for the source, $N_{\rm src}$, above a certain neutrino energy $E_\nu$ for ten years of running 
of the KM3NeT detector.}
\end{table}

%%%%%%%%%%%%%%%%%%%%%%%%%%%%%%%%%%%%%%%%%%%%%%%%%%%%%%%%%%%%%%%%%%%%%%
\section{\label{sec:conclusion} Conclusions }
%%%%%%%%%%%%%%%%%%%%%%%%%%%%%%%%%%%%%%%%%%%%%%%%%%%%%%%%%%%%%%%%%%%%%%

In this paper, we have analysed the source eHWC J1825-134. 
In particular, using updated information on the spectrum provided by 
HAWC, we have calculated 
the number of events expected at the KM3NeT detector for 10 years running 
time. 

The source MGRO J1908+06 was predicted to be one of the most promising 
source to be detected at the IceCube detector~\cite{Halzen:2008zj,Kappes:2009zza,GonzalezGarcia:2009jc,Gonzalez-Garcia:2013iha,Halzen:2016seh}. 
Since the source eHWC J1825-134 is one of the most luminous source in the 
southern sky, we have estimated its discovery potential at the KM3NeT detector assuming that its emission is hadronic.  Will eHWC J1825-134 be the first PeVatron source detected by the KM3NeT detector? 

Note that also the BAIKAL-GVD detector~\cite{Avrorin:2019dli,Avrorin:2019ita} in Baikal Lake will have the discovery potential for this source similar to the KM3NeT detector.  We didn't carry out 
an estimation for this detector since its effective area for muon tracks is currently not  public. 

Also the key difference of KM3 detectors in water respect to the IceCube detector is a much better angular resolution of cascade events, which is about 2 degrees instead of about 10 degrees. 
Since the HAWC source eHWC~J1825-134 has degree scale itself, additional signal from this source will come from the cascade channel in water KM3 detectors, besides the tracks events. 
This should increase additionally the sensitivity of KM3 detectors to this source. 

We want to add that a possible combined analysis between tracks and cascades or between data coming from different KM3 detectors could improve the 
sensitivity to this source. 

%%%%%%%%%%%%%%%%%%%%%%%%%%%%%%

\clearpage 

\section*{Acknowledgments}
This project has received funding from the European Union's Horizon 2020 research and innovation programme under the Marie Sk\l{}odowska-Curie grant agreement 
No. 843418. 
SG acknowledges support
from Agence Nationale de la Recherche (grant ANR- 17-CE31-0014) and from the Observatory of Paris (Action Fed\'eratrice CTA).

%%%%%%%%%%%%%%%%%%%%%%%%%%%%%%%%%%%%%%%%%%%%%%%%%%%%%%%%%%%%%%%%%%%%%%
\section*{References}
\bibliographystyle{elsarticle-num.bst}
\bibliography{biblio}

%%%%%%%%%%%%%%%%%%%%%%%%%%%%%%%%%%%%%%%%%%%%%%%%%%%%%%%%%%%%%%%%%%%%%%

\end{document}